# Discovery of a Three-dimensional Topological Dirac Semimetal, Na$_3$Bi

**(Submitted on Aug 22, 2013)**

Z. K. Liu[1*], B. Zhou[2,3*], Z. J. Wang[4], H. M. Weng[4], D. Prabhakaran[2], S. -K. Mo[3], Y. Zhang[3],

Z. X. Shen[1], Z. Fang[4], X. Dai[4], Z. Hussain[3], and Y. L. Chen[2,3§]

[1.] *Stanford Institute for Materials and Energy Sciences, SLAC National Accelerator Laboratory, 2575 Sand Hill Road, Menlo Park, California 94025*

[2.] *Physics Department, Clarendon Laboratory, University of Oxford, Parks Road, OX1 3PU, UK*

[3] *Advanced Light Source, Lawrence Berkeley National Laboratory, Berkeley California, 94720, USA.*

[4.] *Beijing National Laboratory for Condensed Matter Physics and Institute of Physics, Chinese Academy of Sciences, Beijing 100190, China.*

*: These authors contributed equally to this work
§: Correspondence should be addressed to: Yulin.Chen@physics.ox.ac.uk

**Three-dimensional (3D) topological Dirac semimetals (TDSs) represent a novel state of quantum matter that can be viewed as "3D graphene". In contrast to two-dimensional (2D) Dirac fermions in graphene or on the surface of 3D topological insulators, TDSs possess 3D Dirac fermions in the bulk. The TDS is also an important boundary state mediating numerous novel quantum states, such as topological insulators, Weyl semi-metals, Axion insulators and topological superconductors. By investigating the electronic structure of Na$_3$Bi with angle resolved photoemission spectroscopy, we discovered 3D Dirac fermions with linear dispersions along all momentum directions for the first time. Furthermore, we demonstrated that the 3D Dirac fermions in Na$_3$Bi were protected by the bulk crystal symmetry. Our results establish that Na$_3$Bi is the first model system of 3D TDSs, which can also serve as an ideal platform for the systematic study of quantum phase transitions between rich novel topological quantum states.**

In the past few years, the discoveries of graphene and topological insulators (TIs) have sparked enormous interest in the research of Dirac and topological quantum materials (*1-4*). Graphene is a single sheet of carbon atoms that possess 2D Dirac fermions in its electronic structure (*1, 2, 5*); and TIs are materials with bulk energy gap but gapless surface states formed by an odd number of Dirac fermions with helical spin texture (*3, 4, 6-8*). With the swift development in both fields, questions, such as whether there exists a 3D counterpart of graphene, and whether there exist different materials (other than insulators) that can possess unusual topology in their electronic structures, were naturally raised.

Interestingly, the answers to both questions can lie in a same type of novel quantum matter – the topological Dirac semimetal. In a topological Dirac semimetal, the bulk conduction and valance bands contact only at discrete (Dirac) points and disperse linearly along all (three) momentum directions, forming bulk (3D) Dirac fermions – this naturally makes it the 3D counterpart of graphene. Remarkably, although the point-touching electronic structure has been discussed since more than seven decades ago (*9*), its topological classification was only appreciated recently (*10-15*), leading to the theoretical proposal of topological Dirac semimetals (*12-15*).

The distinct electronic structure of a topological Dirac semimetal not only makes it possible to realize the many exciting phenomena and applications of graphene (*16*) in 3D materials, but also gives rise to many unique properties, such as the giant diamagnetism that diverges logarithmically when the Fermi-energy ($E_F$) is approaching the 3D Dirac point (*17, 18*), quantum magnetoresistance showing linear field dependence in the bulk (*19, 20*), unique Landau level structures under strong magnetic field and oscillating quantum spin Hall effect in its quantum well structure (*15, 21*). Interestingly, the 3D Dirac fermion in a topological Dirac semimetal is composed of two overlapping Weyl fermions (which are chiral massless particles previously studied extensively in high-energy physics, e.g. as a description of neutrinos) (*22*) that can be separated in the momentum space - if time reversal or inversion symmetry is broken - to form the topological Weyl semimetal, another novel topological quantum state showing unique Fermi-arcs geometry (*12, 23*), exhibiting pressure induced anomalous Hall effect (*24*) and quantized anomalous Hall effect in its quantum well structure (*23*).

Besides these unusual properties, the topological Dirac semimetal can be regarded as the boundary state that mediating many other novel states (Fig. 1A, B) (*12-15, 25*) ranging from regular band insulators to topological superconductors – this versatility makes the topological Dirac semimetal an ideal mother compound for the realization of other novel states as well as a platform for the systematic study of topological quantum phase transitions.

Unfortunately, despite these strong motivations, the realization of the topological Dirac semimetal is challenging. In principle, it may be accidentally realized through topological phase transitions, such as tuning chemical composition or spin-orbital coupling strength to the quantum critical point through a normal insulator - topological insulator transition (Fig. 1B). However, such accidental realization is fragile, as the exact chemical composition is hard to control and sensitive to synthesis conditions; and the spin-orbital coupling strength in different materials is not a continuously tunable parameter. Recently, it was realized that the crystal symmetry may protect and stabilize 3D Dirac points and several stoichiometric compounds have been theoretically proposed as topological Dirac semimetals, such as β-cristobalite $BiO_2$ (*14*) and $A_3Bi$ (*A*=Na, K, Rb) family of compounds. (*15*). Due to the metastable nature of β-cristobalite $BiO_2$, we chose $Na_3Bi$ as the subject of study in this work.

We performed angle resolved photoemission spectroscopy (ARPES) measurements to investigate the electronic structures of $Na_3Bi$ (001) single crystals. Further details of the sample preparations and ARPES experiments are available in the supporting online material (*26*). By studying the band dispersions along all momentum directions in the reciprocal space ($k_x$, $k_y$ and $k_z$), we discovered that $Na_3Bi$ possesses a pair of 3D Dirac fermions near the Γ point, with weak in-plane ($k_x$-$k_y$) but strong out-of-plane ($k_z$) anisotropy. Furthermore, we observed that these bulk

Dirac fermions were robust against direct *in-situ* surface modification, indicating the protection of the bulk crystal symmetry. These findings established that $Na_3Bi$ is the first 3D topological Dirac semimetal discovered with stable bulk Dirac fermions.

The crystal structure of $Na_3Bi$ is shown in Fig. 1C, which comprises of stacking …Na - (Na/Bi) - Na… triple-layer groups, with the adjacent triple-layers rotated by 60 degrees with respect to each other. The 3D Brillouin zone (BZ) of $Na_3Bi$ is illustrated in Fig. 1D with high symmetry points indicated. According to recent *ab initio* calculations (*15*), the pair of 3D Dirac fermions locate near the $\Gamma$ point in each BZ (labeled as "D" in Fig. 1D), with linear dispersion along $k_x$, $k_y$, and $k_z$ directions. Since a 3D Dirac fermion is a surface in four-dimensional space ($E = V_x \cdot \vec{k_x} + V_y \cdot \vec{k_y} + V_z \cdot \vec{k_z}$), in Fig. 1E, we visualize it by using $k_x$, $k_y$ and $k_z$ as variables and colors as representation of $E$ (the 4$^{th}$ dimension). On the other hand, if we project the 3D Dirac fermion as two 2D Dirac fermions, they will be much easier to illustrate, as shown in Fig 1F, G. Interestingly, our calculations indicate that while the bulk Dirac cone has small in-plane anisotropy ($V_x \approx V_y = 3.74 \times 10^5$ m/s), the out-of-plane anisotropy is significant ($V_z = 2.89 \times 10^4$ m/s), as demonstrated in Fig.1G.

The overall electronic structure from ARPES measurements is summarized in Fig. 2, from which the characteristic peaks of Na and Bi elements are evident in the core-level spectra (Fig. 2A); and the band dispersions (Fig. 2B) agree well with our *ab initio* calculations [details of the calculations can be found in (*26*)]. The electronic structure at the Dirac point is illustrated in Fig. 2C, showing a cone-shape with linear dispersions [details on selecting $k_z$ positions and identifying Dirac points in momentum space can be found in (*26*)]. This Dirac cone dispersion results in a point-like Fermi-surface (FS) at $\bar{\Gamma}$ in the projected 2D BZ (Fig. 2D). Moreover, the constant energy contours of the Dirac cone at different binding energies (Fig. 2E) demonstrate the small in-plane anisotropy.

To demonstrate the 3D nature of the Dirac cone in Na$_3$Bi, we need to illustrate that the band dispersion is also linear along $k_z$ (as well as $k_x$ and $k_y$), which was achieved by performing photon-energy dependent ARPES measurements [whose principle and details can be found in (*26, 27*)]. By assembling the measurements under broad photon energies (*26*), we obtained the whole band structures of Na$_3$Bi throughout the entire 3D BZ. As an example, Fig. 3A illustrates the complete FS of Na$_3$Bi in a 3D BZ, showing a pair of point-like FSs in the vicinity of Γ ($k_x=k_y=0$ and $k_{z\pm} = \pm 0.08$ *1/Å* or $\pm 0.25\pi/c$, where *c* is the z-direction lattice constant), agreeing well with our *ab initio* calculations which predicts two Dirac points at $k_{z\pm} = \pm 0.26\pi/c$). Note that in Fig. 3A, the additional cylindrical FS vertically crossing the whole 3D BZ (thus is dispersionless along $k_z$) is originated from the surface states (*26, 27*) that emerge during the ARPES measurements with the loss of the surface $N_a$ atoms [which easily migrate away - details can be found in the online supporting materials (*26*)].

Besides the 3D FS, we can investigate the band dispersion along all three *k*-directions. As shown in the schematics of Fig. 3B-D: for a 3D Dirac fermion, ARPES measurements along any *k*-direction should yield either linear or hyperbolic dispersions depending on whether the measurement cuts through the Dirac point – this is different from the usual parabolic dispersions of massive electrons.

In Fig. 3B-D (i-iii), we use three examples along each *k*-direction to show typical ARPES dispersions on Na$_3$Bi [more measurements and analysis can be found in the online supporting materials (*26*)]. Evidently, along each *k*-direction, the dispersion evolves from linear [Fig. B-D (i)] to hyperbolic [Fig. B-D (ii, iii)] shape as expected. Also remarkably is that to fit all of the ARPES measurements [including those in (*26*)], we only need one set of 3D Dirac cone parameters ($V_x$=2.75 eV•A or 4.17×10$^5$m/s, $V_y$=2.39 eV•A or 3.63×10$^5$m/s, and $V_z$=0.7 eV•A or 1.1×10$^5$m/s). These excellent agreements unambiguously prove that the bulk band structure of Na$_3$Bi forms 3D

Dirac cones. Finally, our measurements also confirmed the large anisotropy along the $k_z$ direction ($V_z \approx 0.27 V_x$, see Fig. 3 E, F).

To verify that the 3D Dirac fermion in Na$_3$Bi is protected by the bulk crystal symmetry, we deliberately modified the sample surface (by *in-situ* evaporating *K*-atoms, see Fig. 4A, B) and monitored the band dispersions' evolution with the increase of surface impurities. In addition, the *K*-doping can compensate the charge loss due to the surface $N_a$-atom loss discussed above [and in (*26*)]. In fact, we could even over-compensate the charge loss with enough dosage and observed the upper part of Dirac cone beyond the Dirac point (Fig. 4C). During the *K*-doping, the $E_F$ upshifts (Fig. 4D-F) with the diminishing (Fig.4E, F) of the surface state dispersion (due to the deterioration of the sample surface by randomly deposited *K*-atoms). On the other hand, the bulk Dirac cone (both the linear dispersion and the Dirac point) persists despite such surface deterioration (Fig. 4E, F), supporting that the Dirac fermion is protected by the bulk crystal symmetry.

The discovery of the topological Dirac semimetal Na$_3$Bi realizes the first 3D counterpart of graphene, thus opens the door to realize many exciting applications of graphene in 3D materials. Furthermore, the extremely long Fermi-wavelength (which diverges at the Dirac points) of the bulk conducting electrons in a topological Dirac semimetal can greatly enhance the Ruderman-Kittel-Kasuya-Yosida (RKKY) interaction, making it possible to realize ferromagnetic states by unusually dilute magnetic doping [similar to the Dirac surface states in a topological insulator (*28*)] – which could make the topological Dirac semimetal an ideal platform for spintronics applications.

24. K.-Y. Yang, Y.-M. Lu, Y. Ran, *Phys. Rev. B,* **84**, 075129 (2011).

25. F. R. Klinkhamer, G. E. Volovik, *Int. J. Mod. Phys. A,* **20**, 2795 (2005).

26. Materials and methods are available as supporting online materials.

27. Y. Chen, *Front. Phys.,* **7**, 175 (2012).

28. Q. Liu, C.-X. Liu, C. Xu, X.-L. Qi, S.-C. Zhang, *Phys. Rev. Lett.,* **102**, 156603 (2009).

29. We thank X. L. Qi and Z. Wang for insightful discussions. Y. L. C and B. Z, acknowledge the support from the EPSRC (UK) grant EP/K04074X/1 and a DARPA (US) MESO project (no. N66001-11-1-4105). Z. K. L and Z. X. S acknowledge the support from the Department of Energy under contract DE-AC02-76SF00515. Z.F and X.D acknowledge the supports by the NSF of China, the National Basic Research Program of China, and the International Science and Technology Cooperation Program of China.
Page | 8

**Figure 1:** (**A**) Schematic shows the topological Dirac semimetal (TDS) state and various neighboring states. (**B**) Illustration shows the realization of the TDS state at the quantum critical point in the topological quantum phase transition from a normal insulator to a topological insulator. The "+" and "–" signs denote the even and odd parity of the bands at the time reversal invariant point, respectively. (**C**) Crystal structure and (**D**) the Brillouin zone (BZ) of $Na_3Bi$. Cyan dots indicate the high symmetry points of the BZ; and the red dots highlight the 3D Dirac point positions. (**E**) Visualization of a 3D Dirac fermion dispersion ($E = V_x \cdot \vec{k_x} + V_y \cdot \vec{k_y} + V_z \cdot \vec{k_z}$), and different color is used to represent energy. (**F, G**) Projection of the 3D Dirac fermion onto ($k_x$, $k_y$, $E$) and ($k_x$, $k_z$, $E$) spaces (see text). Red lines outline the linear dispersions along $k_x$, $k_y$ and $k_z$ directions.

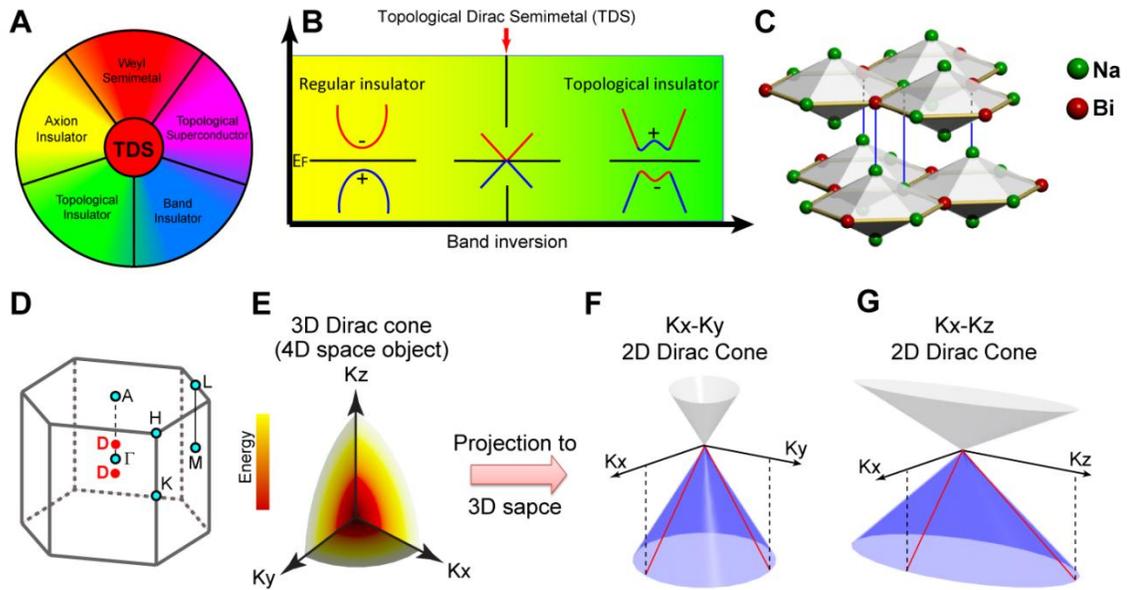

**Figure 2:** General electronic structures of Na$_3$Bi (**A**) Core level photoemission spectrum clearly shows strong characteristic Na *2p* and Bi *5d* doublet peaks (blue curve). Features at low binding energy (E$_B$< 21eV) were magnified 20 times to enhance the details (red curve). (**B**) Comparison of the valance band spectra between ARPES measurement (background) and *ab initio* calculations (solid lines) along the $\bar{M} - \bar{\Gamma} - \bar{M}$ direction. The color of the lines represents different *k$_z$* dispersions in the calculation. (**C**) 3D intensity plot of the photoemission spectra at the Dirac point, showing cone-shape dispersion. (**D**) Broad FS map from ARPES measurements that covers there BZs. The red hexagons represent the surface BZ; and the uneven intensity of the FS points of different BZ results from the matrix element effect. (**E**) Stacking plot of constant energy contours at different binding energy shows Dirac cone band structure. Red dotted lines are the guides to the eye that indicate the dispersions and intersect at the Dirac point.

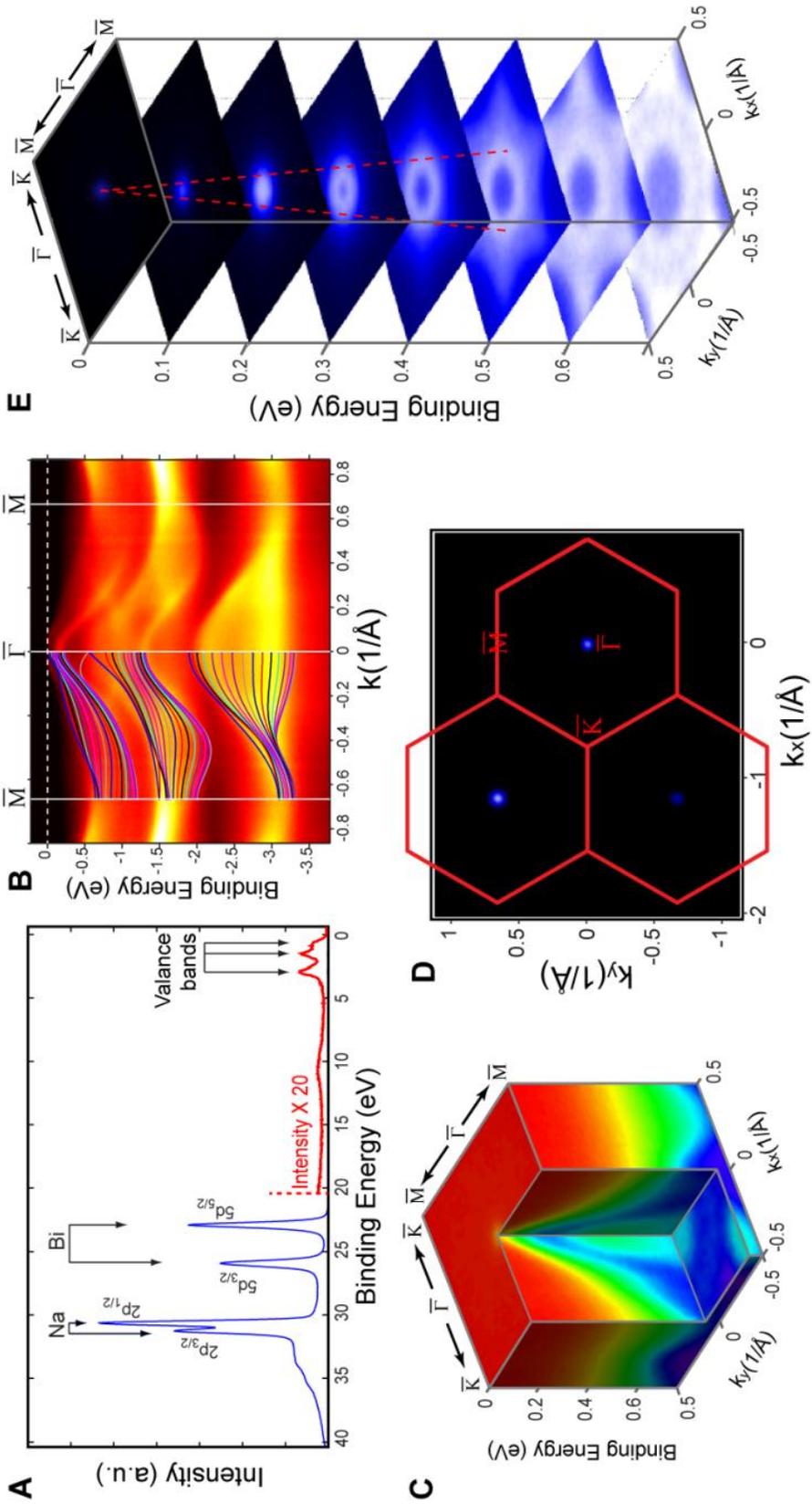

**Figure 3:** Dispersion of the 3D Dirac fermion along all three momentum directions. (**A**) FS map across the whole 3D BZ (top panel) and its projection to the surface BZ (bottom panel). Compared to Fig. 2**D**, the additional circular FS around the $\bar{\Gamma}$ point (bottom) and the cylindrical FS (top panel) result from the surface state due to the migration of the active surface Na atoms (see text and [28] for detailed discussions). (**B-D**) Schematics show the dispersions (red curves) obtained by ARPES measurements that slice through the 3D Dirac cones at different ($k_x$, $k_y$ or $k_z$) momentum locations, showing either linear or hyperbolic shape. **B**(i-iii) Measured dispersion at $k_x$=0, 0.06 Å$^{-1}$ and 0.11 Å$^{-1}$, respectively. Red dotted lines shows the fitted dispersions that agree well with the experiments (see text and [28] for details). **C**(i-iii) Experiment and fitted dispersions at $k_y$=0, 0.06 Å$^{-1}$ and 0.11Å$^{-1}$, respectively. **D**(i-iii) Experiment and fitted dispersions at $k_z$=0, 0.13 Å$^{-1}$ and 0.24 Å$^{-1}$, respectively. (**E-F**) The two projected Dirac cones reconstructed by the experiment parameters: $V_x$=2.75 eV•A, $V_y$=2.39 eV•A, and $V_z$=0.70 eV•A [see (26) for details].

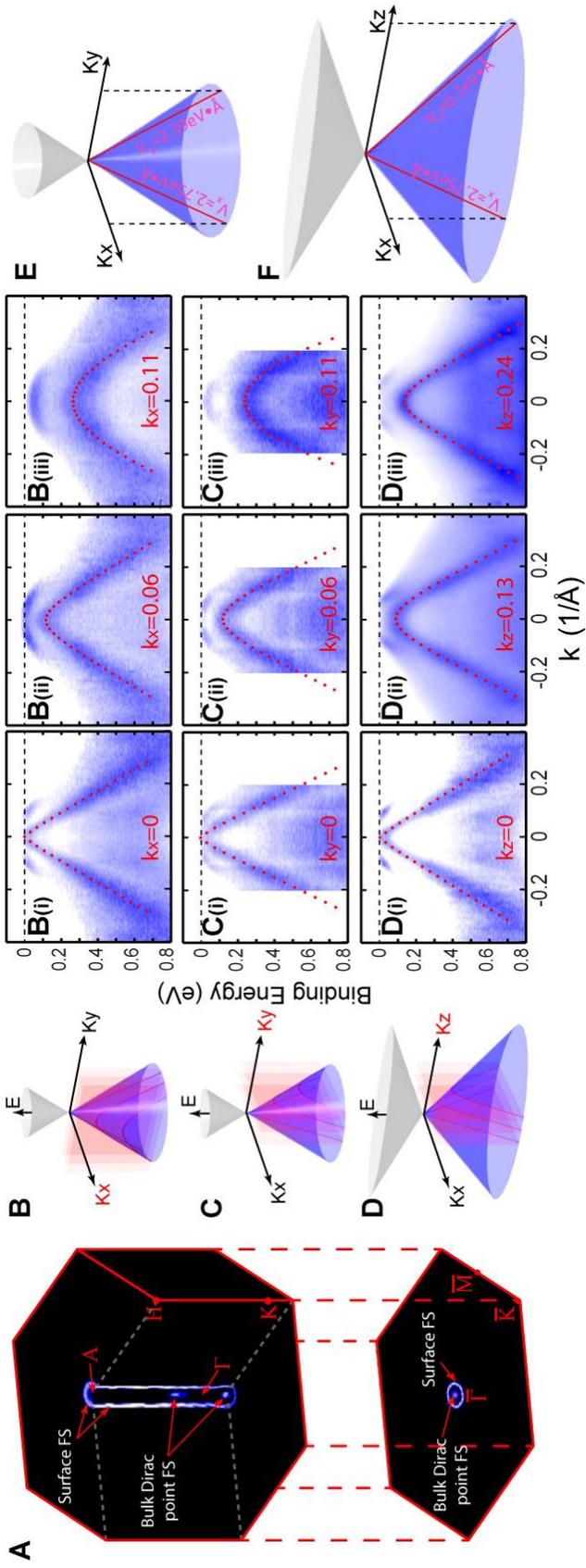

**Figure 4:** Test of bulk Dirac fermion stability and $E_F$ tuning by surface doping. (**A**) Illustration of the *in-situ* surface *K*-doping. (**B**) Core level photoemission spectra before and after the *K*-doping show the rise of the $K_{3p}$ core level peak (which can be used to monitor the dosage level). (**C**) Stacking plot of constant energy contours shows the upper Dirac cone after the *in-situ K*-doping. White dashed lines are the guides to the eye that trace the Dirac dispersions. (**D-F**) ARPES intensity plots show the rising of the $E_F$ position with *K*-dosage. (**D**): Before *in-situ K*-doping; (**E**) precise *K*-doping to bring $E_F$ to the bulk Dirac point; (**F**): further *K*-doping drive the system into n-type. Note the surface state band (SSB) in (**D**) is destroyed by the *K*-doping, thus does not show in (**E**) and (**F**).

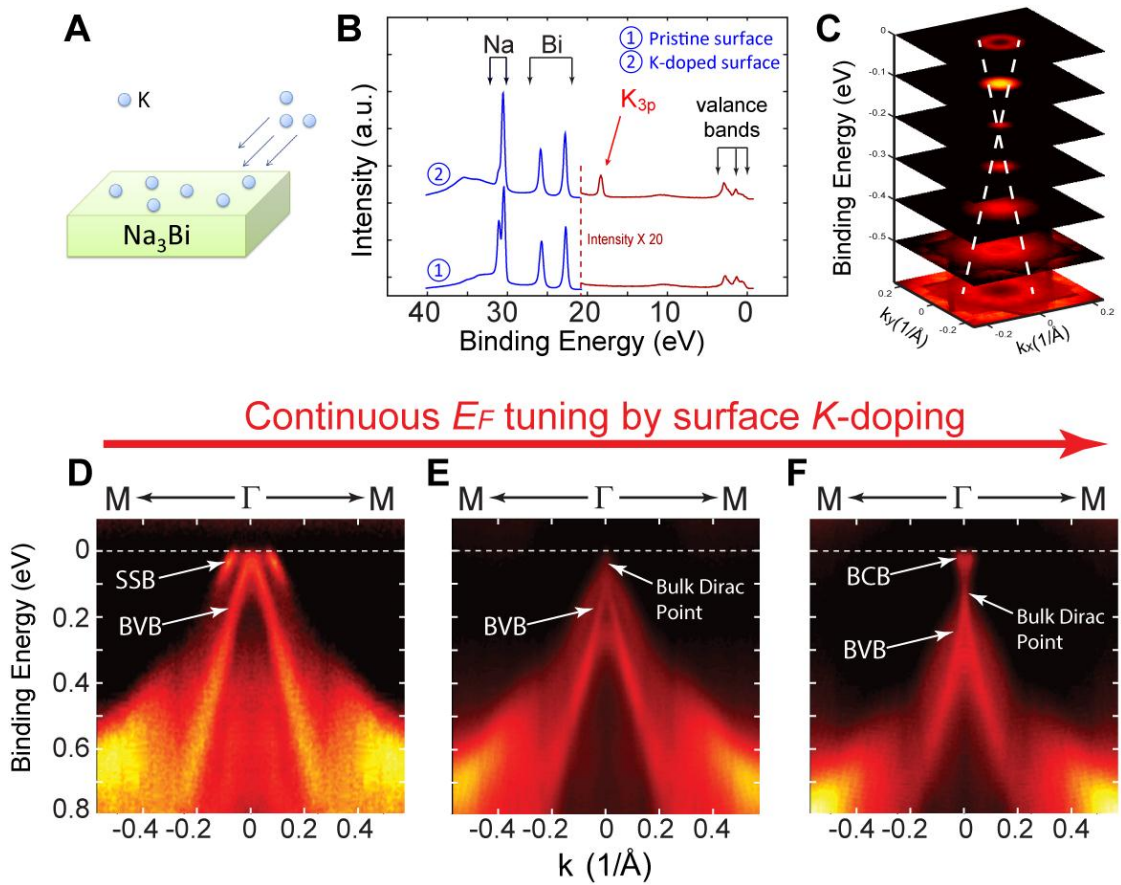

# Supporting Online Materials

# SOM A: Materials and Methods:

*A1*: **Sample growth and preparations:**

High quality $Na_3Bi$ single crystals were synthesized by using stoichiometric amounts of high purity elements of Na (99.99%) and Bi (99.999%), which were put in a Tantalum tube under Argon atmosphere and arc-welded. The sealed Ta-tube was then placed inside a quartz tube and sealed under vacuum. The quartz tube was placed in a vertical tube furnace and heated to 950 °C for 6 hours before cooling down to 750 °C at the rate of 2 °C/hour; after which it was cooled down to room temperature at the rate of 60 °C/hour.

To minimize the degradation of the sample by chemical reactions with atmosphere, the samples were handled in Argon-filled glove box after growth; stored and transported in sealed ampules; then mounted in Argon-filled glove box in-site for the ARPES experiments.

*A2*: **Angle resolved photoemission spectroscopy (ARPES) measurements**

ARPES measurements were performed at beamline 10.0.1 of Advanced Light Source (ALS) at Lawrence Berkeley National Laboratory. The measurement pressure was kept $< 3 \times 10^{-11}$ Torr, and data were recorded by a Scienta R4000 analyzer at 100 K sample temperature. The total convolved energy and angle resolutions were 15meV and 0.2° (i.e. $0.01 Å^{-1} \sim 0.015 Å^{-1}$ for photoelectrons generated by $41eV \sim 75eV$ photons used in our experiments)

The fresh surfaces for ARPES measurement were obtained by cleaving the $Na_3Bi$ samples *in-situ* along its natural cleavage plane (001). We noticed that the samples became slightly hole doped shortly after cleaving process (typically within 30~60 minutes), due to the missing of surface Na atoms (as alkaline atoms can easily migrate away from the surface in UHV environment); the doping level then became stable during the ARPES measurements (typical measurement duration was ~ 24 hours).

To compensate the charge loss due to the surface Na-deficiency discussed above, alkaline metal atoms (e.g. $N_a$, $K$, $C_s$) can be introduced to the sample surface. Section **SOM C** below will detail such *in-situ* surface *K*-doping in our ARPES measurements.

*A3*: ***Ab initio* calculations of the band structures:**

We employed the projector augmented wave method [S1, S2] implemented in VASP [S3] within generalized gradient approximation (GGA) with Perdew-Burke-Ernzerhof type functional [S4] for the exchange-correlation potential. The cut-off energy for the plane wave basis was 325 eV, and a 12x12x6 k-grid was used for sampling the Brillouin zone. The convergence of these settings has been checked. As shown in Fig. 1C in the main text, $Na_3Bi$ has layered structure stacking along (001) direction, and a supercell with 10 layers was used for the (001) surface state calculation.

# SOM *B*: Photon energy dependent ARPES measurements

*B1*: **$k_z$-momentum determination:**

In an ARPES measurement, the in-plane electron momentum ($k_{//}$, parallel to the sample surface) can be naturally determined by the momentum conservation of photoelectrons [S5]; while determining the out-of-plane momentum component ($k_z$) is less straightforward - which requires a series ARPES measurements performed under different photon energies [S5, S6].

Based on the free-electron final state approximation with a potential parameter $V_0$ (also known as the inner potential) describing the energy difference of photoelectrons before and after leaving the crystal surface, we can derive the $k_z$ as [S6, S7]:

$$k_z = \frac{\sqrt{2m_e(E_k \cos^2\theta + V_0)}}{\hbar}$$

where θ is the emission angle and $E_k$ is the kinetic energy of the emitted electron, which satisfies:

$$E_k = h\upsilon - w - E_B$$

where $h\upsilon$ is the photon energy, $w$ is the work function of the sample and $E_B$ is the electron binding energy (see Fig. S1).

As $V_0$ varies with samples, we performed energy dependent ARPES by using a broad range of photon energies to cover enough $k_z$–span (ideally more than the $k_z$ size of one BZ), and used the high symmetry points in the BZ to identify the exact value of $V_0$.

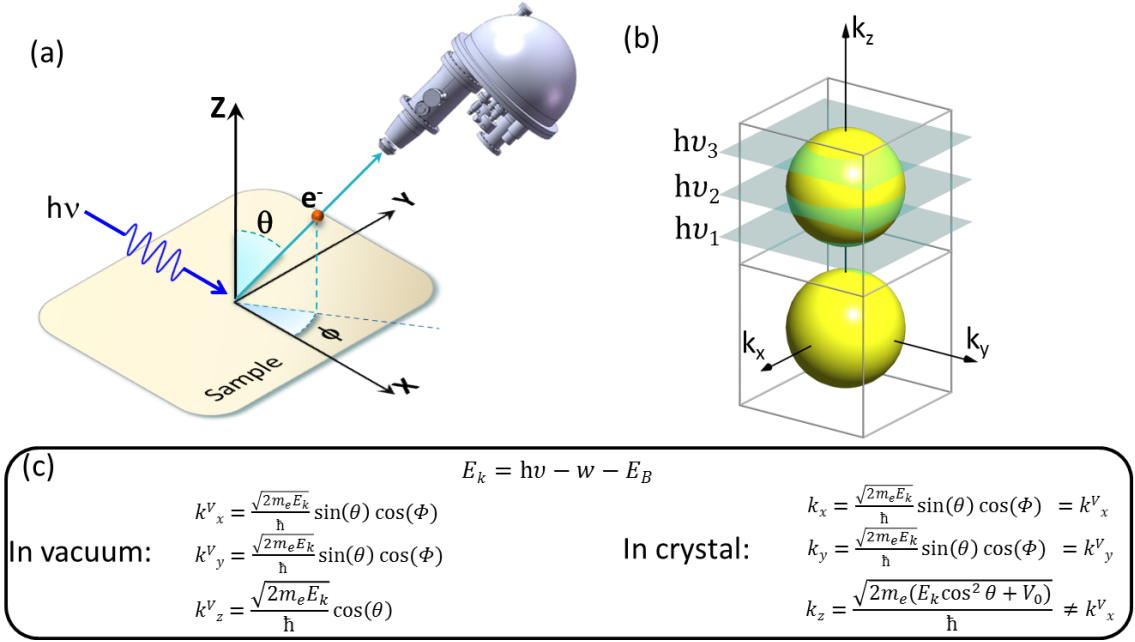

**Fig. S1. Illustration of $k_z$ dependent ARPES measurements by varying photon energy.** (a) Schematic showing the principle of an ARPES measurement. (b) A schematic showing that by using different photon energy, electronic structures at different $k_z$ in a BZ can be probed. (c) Information we can extract from a photoelectron: (1) the photoelectron's energy and momenta along all three directions can be easily deduced; and we can also deduce (2) the energy and momenta of the electron before photo-emitted – please note the difference in the $k_z$ components for the electron before (left) and after (right) the photoemission process.

*B2*:  **Complete band and Fermi-surface mapping throughout an entire 3D Brillouin zone**

Making use of the broad photon energy range of Beamline 10.0.1 at ALS, we performed photon energy dependent ARPES (see Fig. S1 above) to obtain the complete electronic structures of $Na_3Bi$ (band dispersions and Fermi-surfaces) throughout an entire 3D BZ.

We scanned the photon energy from 41eV to 75eV in the ARPES measurements (which covered ~1.3 BZ size along the $k_z$ direction), which enabled us to determine the exact $k_z$ location as well as the inner potential $V_0$ (17eV) discussed above in **SOM *B1***.

After determining $k_z$ (by using $V_0$), together with the in-plane momentum $k_x$, $k_y$ (Fig. S1) and electron energy E, we obtained the **full** electronic structure of $Na_3Bi$ (with all four parameters: $k_x$, $k_y$, $k_z$, and *E*), as illustrated in the main text Fig. 2, 3, and Fig. S6 below.

## SOM *C*:    Surface charge loss and compensation

*C1*:    **Loss of surface charge by Na migration**

Due to the high mobility of alkaline atoms, $N_a$ atoms on the cleaved sample surface can migrate away during the experiments, causing slight hole-doping effect. In Fig. S2, we show the band structure measured ~2 hours after cleaving the sample, which shows clearly hole-doping compared to Fig. 2 **C-E** in the main text (which were obtained within 20 minutes after cleave), we notice that after 1~2 hours, the hole doping level became stable and stopped increasing throughout the rest of our measurements (typically ~24 hours).

As can be seen in Fig. S2, from both the 3D band structure (Fig. S1a) and the dispersions along high symmetry points (Fig. S2b, c), the $E_F$ position resides closely to the top of the bulk valance band, indicating that the sample is slightly p-doped (caused by the $N_a$-atom migration).

Accompanied by this p-doping, we observed an additional sharp dispersion along the external envelope of the bulk valance band (also seen in Fig. 3A in the main text), which forms a large hole-like FS around the 3D Dirac point (Fig. 3A in the main text and Fig. S1a). The surface nature of this band was proved by the fact that this band has NO $k_z$ dispersion - which can be seen in the 3D FS map in the main text Fig. 3A. Also, due to its surface nature, it may be destroyed by surface contamination, as is seen in Fig. 4 in the main text and section **SOM *C2*** below.

As an independent check, we performed *ab initio* calculation, which confirmed that a surface state can emerge along the envelope of the bulk valance band [Fig. S1d and see Ref [S8]].

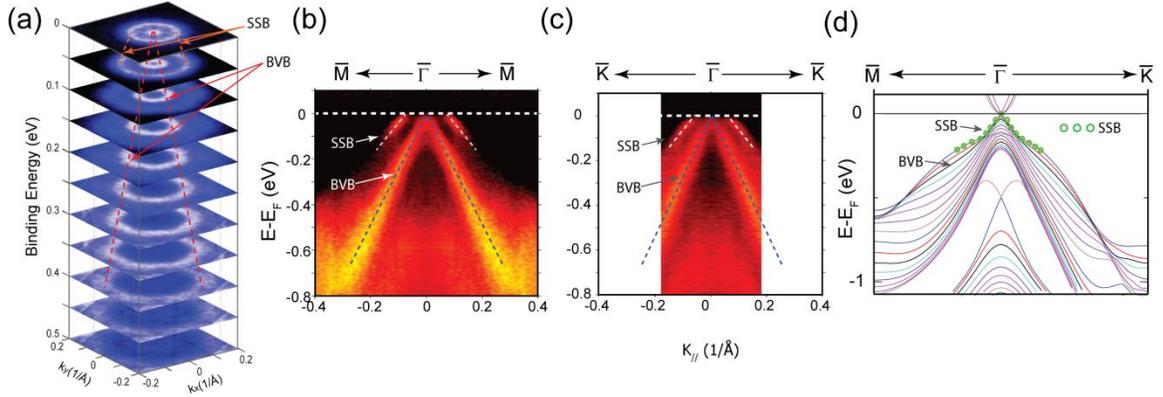

**Fig. S2. Electronic structure of Na$_3$Bi with $N_a$ vacancies from ARPES experiments.** (a) Stacking plot of constant energy contours. In addition to the Dirac cone of the bulk band, another (surface state) hole-pocket is observed. (b, c) Intensity plots of ARPES measurements along high-symmetry directions (Γ-M and Γ-K, respectively) that cut through the 3D Dirac point (at $k_z$=-0.07 Å$^{-1}$). Dotted lines mark the dispersion of the bulk valance band (BVB) and the surface state band (SSB). (d) *Ab initio* calculation shows the surface state band (SSB) along the envelope of the bulk valance band (BVB), agreeing with our measurements.

*C2*: **Compensation of surface charge by *in-situ* K-doping**

To compensate the surface charge loss caused by $N_a$ migration and to double-check the surface nature of the extra band (along the envelope of the bulk band) seen in Fig. S2, we evaporated *K*-atoms onto the sample surface *in-situ* and monitored the surface band evolution by ARPES.

The K-doping effectively raises the $E_F$ position as expected (discussed in the main text Fig. 4), and Fig. S3 below shows more of such evolution measurements around the *Γ* and *A* points in the 3D BZ.

Also evidently, in both evolutions, the surface state dispersions were diminished with the increase of the *K*-impurities on the surface (as they are randomly distributed); while the bulk dispersion remains (note that the *K*-impurities on the surface can also scatter the bulk photoelectrons thus broaden the bulk band spectra).

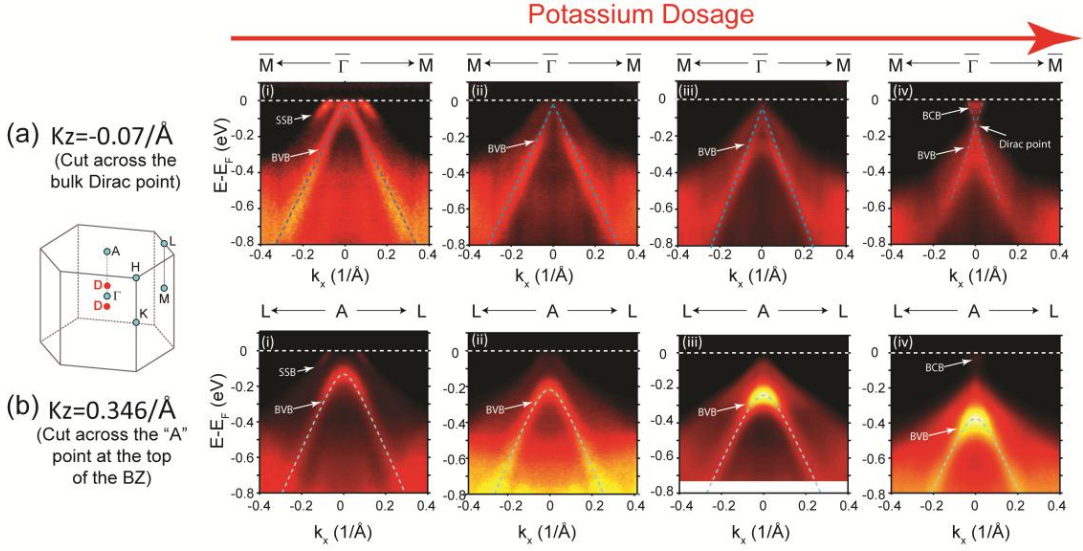

**Fig. S3. Evolution of the electronic structure with surface *K*-doping.** Inset: Schematic of the 3D BZ of $Na_3Bi$. (a) Intensity plot of the evolution of the dispersion along the $\bar{M}$-$\bar{\Gamma}$-$\bar{M}$ direction that cuts through the bulk Dirac point ($k_z$=-0.07Å$^{-1}$) with *K*-dosage. The rising of the $E_F$ position and the diminishing of the surface state band (SSB) is obvious. (b) Same as (a), for L-A-L direction dispersions ($k_z$=0.346Å$^{-1}$). Acronyms: BVB: bulk valance band; BCB: bulk conduction band.

### C3: Loss of doped K-atoms after stopping in-situ K-doping

Similar to the surface $N_a$ atoms, the *K* atoms we dope onto the $Na_3Bi$ surface can also migrate away after we stop the doping process (which is another support showing that the slight p-doping on the sample surface was caused by the $N_a$-loss).

As illustrated in Fig. S4, indeed we observed the downshift of $E_F$ (by ~75meV in 2 hours) after stopping the K-doping, similar to the $E_F$ down-shift in Fig. S2.

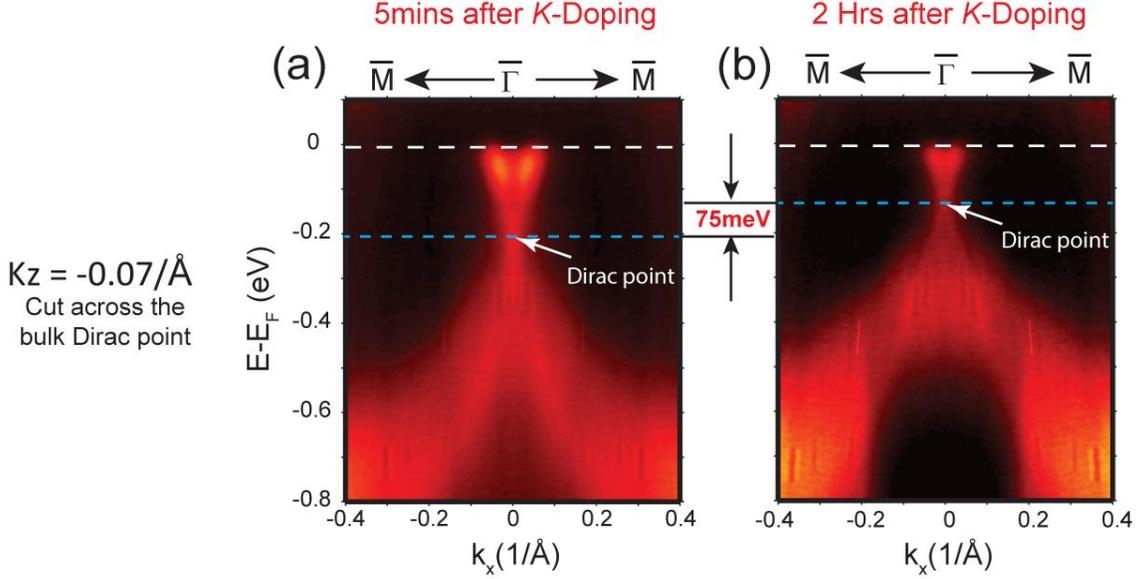

**Fig. S4. Charge loss due to *K*-atom migration after stopping *K*-doping.** (a, b) Intensity plots of the dispersions along $\bar{M}$- $\bar{\Gamma}$-$\bar{M}$ direction that cut through the 3D Dirac point at $k_z$= -0.07 Å$^{-1}$ or -0.25π/c. Spectra in (a) were taken right after (~5 minutes) the surface *K*-doping while those in (b) was taken at ~2 hours after *K*-doping. Comparison between the two measurements reveals a ~75meV downshift of $E_F$ due to the surface *K*-atom loss.

## SOM *D*: 3D Band fitting

### *D1*: Extraction of band dispersions from ARPES measurements

We extracted the band dispersions from ARPES measurements by fitting both the momentum distribution curves (MDCs) and energy distribution curves (EDCs) [S6], and an example is shown in Fig. S5.

For a 3D Dirac cone, the extracted band dispersion in each (2D) ARPES measurement is either linear or hyperbolic (main text Fig. 3B-D), depending on whether the measurement cuts through the 3D Dirac point.

As the linear dispersions in Fig. 3B(i), C(i) and D(i) of the main text are obvious, in Fig. S5 (d, e) we compare the fittings of a hyperbola and a parabola [data from main text Fig. 3D(ii)]. Clearly, the fitting to the hyperbola (Fig. S5d) is excellent while in contrast, the fitting to a parabola (Fig. S5e) shows clear discrepancy.

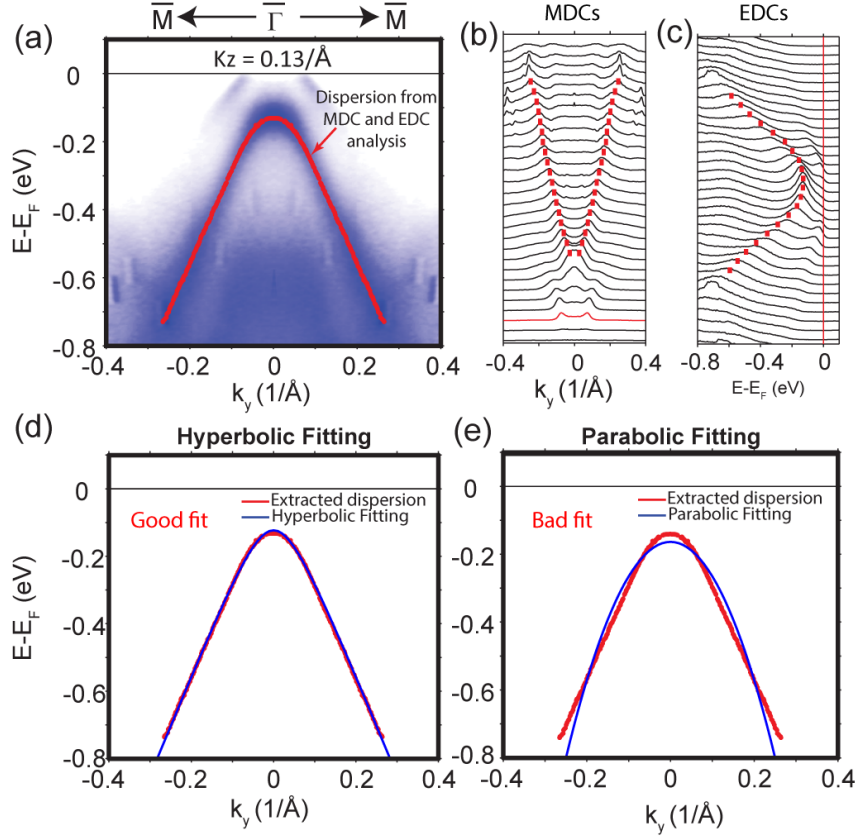

Example:
Fit the data from the main text Fig 3D (ii) by a hyperbola

**Fig. S5. Extracting and fitting the dispersion.** (a) ARPES intensity plot of the measurement along the $\overline{M}$- $\overline{\Gamma}$-$\overline{M}$ direction at $k_z$=0.25 Å$^{-1}$. The red curve indicates the dispersion extracted from fitting the momentum distribution curves (MDCs) and energy distribution curves (EDCs). (b, c) Stack plot of the MDCs (in b) and EDCs (in c) of the spectra in (a). Red squares indicate the fitted peak positions of the MDCs and EDCs, respectively. (d, e) Fitting the dispersion (red curves) from (a) with a hyperbolic (d) and a parabolic curve (e), respectively. Obviously, the fitting in (d) is excellent and that in (e) shows clear discrepancy.

***D2*:    Fit ALL ARPES measurements by ONE set of parameters ($V_x$, $V_y$ and $V_z$).**

Since a 3D Dirac cone can be completely characterized by one set of (velocity) parameters ($V_x$, $V_y$ and $V_z$), we can check if all the dispersions from our ARPES measurements can be fitted with *a single set* of velocity parameters.

As explained in the main text (and part *D1* above), the measured dispersions are either linear or hyperbolic. In addition to the spectra shown in the main text Fig. 3, Fig. S6 illustrates more ARPES measurements along each momentum direction – all of which were well fitted by the same set of velocity parameters used in the main text ($V_x = 2.75$ eV•A, $V_y = 2.39$ eV•A and $V_z = 0.7$ eV•A,).

This global agreement between **_ALL_** the ARPES measurements and the fitting results unambiguously proves that the bulk electronic structure of Na$_3$Bi indeed forms 3D Dirac fermions.

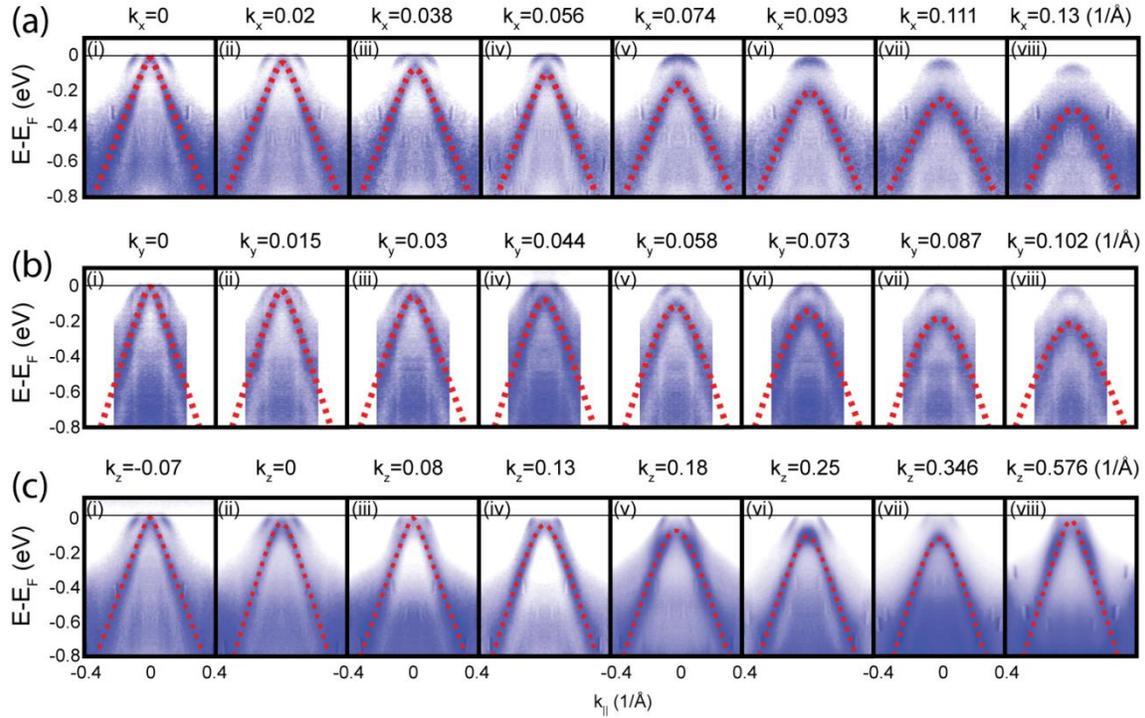

**Fig. S6. Dispersions of 3D Dirac fermions from ARPES measurements with excellent fitted results.** (a-c): Spectra intensity plots from ARPES measurements with different $k_x$, $k_y$ and $k_z$ value, respectively. The measurement geometry is defined in the main text Fig. 3. Red broken lines were fitted dispersions by using only one set of velocity parameters ($V_x = 2.75$ eV•A, $V_y = 2.39$ eV•A and $V_z = 0.7$ eV•A,), showing excellent agreement with the measurements.